
\font\bigbold=cmbx12
\font\eightrm=cmr8
\font\sixrm=cmr6
\font\fiverm=cmr5
\font\eightbf=cmbx8
\font\sixbf=cmbx6
\font\fivebf=cmbx5
\font\eighti=cmmi8  \skewchar\eighti='177
\font\sixi=cmmi6    \skewchar\sixi='177
\font\fivei=cmmi5
\font\eightsy=cmsy8 \skewchar\eightsy='60
\font\sixsy=cmsy6   \skewchar\sixsy='60
\font\fivesy=cmsy5
\font\eightit=cmti8
\font\eightsl=cmsl8
\font\eighttt=cmtt8
\font\tenfrak=eufm10
\font\sevenfrak=eufm7
\font\fivefrak=eufm5
\font\tenbb=msbm10
\font\sevenbb=msbm7
\font\fivebb=msbm5
\font\tensmc=cmcsc10
\font\tencmmib=cmmib10  \skewchar\tencmmib='177
\font\sevencmmib=cmmib10 at 7pt \skewchar\sevencmmib='177
\font\fivecmmib=cmmib10 at 5pt \skewchar\fivecmmib='177

\newfam\bbfam
\textfont\bbfam=\tenbb
\scriptfont\bbfam=\sevenbb
\scriptscriptfont\bbfam=\fivebb

\newfam\frakfam
\textfont\frakfam=\tenfrak
\scriptfont\frakfam=\sevenfrak
\scriptscriptfont\frakfam=\fivefrak

\newfam\cmmibfam
\textfont\cmmibfam=\tencmmib
\scriptfont\cmmibfam=\sevencmmib
\scriptscriptfont\cmmibfam=\fivecmmib


\def\eightpoint{%
\textfont0=\eightrm   \scriptfont0=\sixrm
\scriptscriptfont0=\fiverm  \def\rm{\fam0\eightrm}%
\textfont1=\eighti   \scriptfont1=\sixi
\scriptscriptfont1=\fivei  \def\oldstyle{\fam1\eighti}%
\textfont2=\eightsy   \scriptfont2=\sixsy
\scriptscriptfont2=\fivesy
\textfont\itfam=\eightit  \def\it{\fam\itfam\eightit}%
\textfont\slfam=\eightsl  \def\sl{\fam\slfam\eightsl}%
\textfont\ttfam=\eighttt  \def\tt{\fam\ttfam\eighttt}%
\textfont\bffam=\eightbf   \scriptfont\bffam=\sixbf
\scriptscriptfont\bffam=\fivebf  \def\bf{\fam\bffam\eightbf}%
\abovedisplayskip=9pt plus 2pt minus 6pt
\belowdisplayskip=\abovedisplayskip
\abovedisplayshortskip=0pt plus 2pt
\belowdisplayshortskip=5pt plus2pt minus 3pt
\smallskipamount=2pt plus 1pt minus 1pt
\medskipamount=4pt plus 2pt minus 2pt
\bigskipamount=9pt plus4pt minus 4pt
\setbox\strutbox=\hbox{\vrule height 7pt depth 2pt width 0pt}%
\normalbaselineskip=9pt \normalbaselines
\rm}


\def\pagewidth#1{\hsize= #1}
\def\pageheight#1{\vsize= #1}
\def\hcorrection#1{\advance\hoffset by #1}
\def\vcorrection#1{\advance\voffset by #1}

\newcount\notenumber  \notenumber=1              
\newif\iftitlepage   \titlepagetrue              
\newtoks\titlepagefoot     \titlepagefoot={\hfil}
\newtoks\otherpagesfoot    \otherpagesfoot={\hfil\tenrm\folio\hfil}
\footline={\iftitlepage\the\titlepagefoot\global\titlepagefalse
           \else\the\otherpagesfoot\fi}

\def\abstract#1{{\parindent=30pt\narrower\noindent\eightpoint\openup
2pt #1\par}}
\def\smc{\tensmc}


\def\note#1{\unskip\footnote{$^{\the\notenumber}$}
{\eightpoint\openup 1pt
#1}\global\advance\notenumber by 1}

\def\frac#1#2{{#1\over#2}}

\def\({\left(}
\def\){\right)}
\def\<{\langle}
\def\>{\rangle}
\def\2pd#1#2#3{\frac{\partial^2#1}{\partial#2\partial#3}}

\def\sqr#1#2{{\vcenter{\vbox{\hrule height.#2pt
        \hbox{\vrule width.#2pt height#1pt \kern#1pt
           \vrule width.#2pt}
        \hrule height.#2pt}}}}

\def\ni{\noindent}


\global\newcount\secno \global\secno=0
\global\newcount\meqno \global\meqno=1
\global\newcount\appno \global\appno=0
\newwrite\eqmac
\def\romappno{\ifcase\appno\or A\or B\or C\or D\or E\or F\or G\or H
\or I\or J\or K\or L\or M\or N\or O\or P\or Q\or R\or S\or T\or U\or
V\or W\or X\or Y\or Z\fi}
\def\eqn#1{
        \ifnum\secno>0
            \eqno(\the\secno.\the\meqno)\xdef#1{\the\secno.\the\meqno}
          \else\ifnum\appno>0
            \eqno({\rm\romappno}.\the\meqno)\xdef#1{{\rm\romappno}.\the\meqno}
          \else
            \eqno(\the\meqno)\xdef#1{\the\meqno}
          \fi
        \fi
\global\advance\meqno by1 }

\def\eqnn#1{
        \ifnum\secno>0
            (\the\secno.\the\meqno)\xdef#1{\the\secno.\the\meqno}
          \else\ifnum\appno>0
            \eqno({\rm\romappno}.\the\meqno)\xdef#1{{\rm\romappno}.\the\meqno}
          \else
            (\the\meqno)\xdef#1{\the\meqno}
          \fi
        \fi
\global\advance\meqno by1 }


\pageheight{24cm}
\pagewidth{15.5cm}
\magnification \magstep1
\voffset=8truemm
\baselineskip=16pt
\parskip=5pt plus 1pt minus 1pt


\global\newcount\refno
\global\refno=1 \newwrite\reffile
\newwrite\refmac
\newlinechar=`\^^J
\def\ref#1#2{\the\refno\nref#1{#2}}
\def\nref#1#2{\xdef#1{\the\refno}
\ifnum\refno=1\immediate\openout\reffile=refs.tmp\fi
\immediate\write\reffile{
     \noexpand\item{[\noexpand#1]\ }#2\noexpand\nobreak.}
     \immediate\write\refmac{\def\noexpand#1{\the\refno}}
   \global\advance\refno by1}
\def\semi{;\hfil\noexpand\break ^^J}
\def\nl{\hfil\noexpand\break ^^J}
\def\refn#1#2{\nref#1{#2}}

\def\up#1{$^{[#1]}$}

\def\mplA#1#2#3{{\it Mod. Phys. Lett.} {\bf A{#1}} (19{#2}) #3}
\def\pl#1#2#3{{\it Phys. Lett.} {\bf {#1}B} (19{#2}) #3}

\def\pr#1#2#3{{\it Phys. Rev.} {\bf {#1}} (19{#2}) #3}

\def\prD#1#2#3{{\it Phys. Rev.} {\bf D{#1}} (19{#2}) #3}
\def\prl#1#2#3{{\it Phys. Rev. Lett.} {\bf #1} (19{#2}) #3}

\def\prp#1#2#3{{\it Phys. Rep.} {\bf {#1}} (19{#2}) #3}
\def\tmp#1#2#3{{\it Theor. Math. Phys.} {\bf {#1}} (19{#2}) #3}


\input epsf
\def\fig#1#2{$$\epsfxsize=#2\epsfbox{#1}$$}


\def\pa{\partial}

\def\cd{{\cdot}}
\def\hm{h^{-1}}

\def\intden{\int\!\frac{d^4k}{(2\pi)^4}}
\def\zz#1{Z_2^{#1}}


\secno=0

{
\parskip=2pt
\refn\FAD{L.\ Faddeev and P.\ Kulish, \tmp{4}{70}{745}}

\refn\KIBBLE{T.W.B.\ Kibble, \pr{173}{68}{1527}}

\refn\JR{J.M.\ Jauch and F.\ Rohrlich, {\sl The Theory of
Photons and Electrons}, 2nd expanded edition
(Springer-Verlag, New York 1980)}

\refn\CSS{For a review, see J.C. Collins, D.E. Soper and G. Sterman,
in {\sl Perturbative Quantum Chromodynamics}, ed. A.H. Mueller (World
Scientific, Singapore, 1989)}

\refn\CONT{H.H.\ Contanopanagos and M.B.\ Einhorn, \prD{45}{92}{1291}}

\refn\TAYL{J.C.\ Taylor, \prD{54}{96}{2975}}

\refn\CONTR{H.H.\ Contanopanagos and M.B.\ Einhorn,
\prD{54}{96}{2978}}

\refn\CIAFALONI{M.\ Ciafaloni, in {\sl Perturbative
Quantum Chromodynamics},
edited by A.H.\ Mueller (World Scientific, Singapore 1989)}

\refn\US{M.\ Lavelle and D.\ McMullan, \prp{279}{97}1}

\refn\GEORGI{H.\ Georgi, \pl{240}{90}{447}}

\refn\FIRST{T. Kawai and H.P. Stapp, \prD{52}{95}{2484, 2505, 2517}}

\refn\HALL{L. Chen, M. Belloni and K. Haller, \prD{55}{97}{2347}}

\refn\KASH{T. Kashiwa and N. Tanimura, hep-th/9605207}

\refn\HAJO{P. Haagensen and K.\ Johnson, hep-th/9702204}

\refn\LAST{D. Alba and L. Lusanna, hep-th/9705155}

\refn\FAST{E.\ Bagan, M.\ Lavelle and D.\ McMullan,
\prD{56}{97}{3732}}
\refn\SCAL{E.\ Bagan, B.\ Fiol, M.\ Lavelle and D.\ McMullan,
\mplA{12}{97}{1815}; Erratum ibid. {\bf A12} (1997) 2317}

\refn\JACKSOL{R.\ Jackiw and L.\ Soloviev, \pr{173}{68}{1485}}

\refn\ZWAN{For a recent reference, see:
A.\ Cucchieri and D.\ Zwanziger, \prl{78}{97}{3814}}

}
%
%
\rightline {UAB-FT-417}
\rightline {PLY-MS-97-23}
\vskip 38pt
\centerline{\bigbold Soft Dynamics and Gauge Theories}
\vskip 28pt
\centerline{\smc Emili Bagan,{\hbox {$^1$}}
Martin Lavelle{\hbox {$^2$}} and  David McMullan{\hbox {$^2$}}}
\vskip 15pt
{\baselineskip 12pt \centerline{\null$^1$Grup de F\'\i sica Te\`orica
and IFAE}
\centerline{Edificio Cn}
\centerline{Universitat Aut\`onoma de Barcelona}
\centerline{E-08193 Bellaterra (Barcelona)}
\centerline{Spain}
\centerline{email: bagan@ifae.es}

\vskip 13pt
\centerline{\null$^{2}$School of Mathematics and Statistics}
\centerline{University of Plymouth}
\centerline{Drake Circus, Plymouth, Devon PL4 8AA}
\centerline{U.K.}
\centerline{email: mlavelle@plymouth.ac.uk}
\centerline{email: dmcmullan@plymouth.ac.uk}}
\vskip 35pt
{\baselineskip=13pt\parindent=0.58in\narrower\ni{\bf Abstract}\quad
Infra-red divergences obscure the underlying soft dynamics in gauge
theories. They remove the
pole structures associated with particle propagation in the various
Green's functions of gauge theories.
Here we present a solution to this
problem. We give two equations which describe
how charged particles must be dressed by gauge
degrees of freedom. One follows from
gauge invariance, the other, which is new,
from velocity superselection rules
familiar from the heavy quark effective theory. The
solution to these equations in the abelian theory is proven to
lead to on-shell Green's
functions that are free of soft divergences at all orders in
perturbation theory.
\par}

\vfill\eject
\noindent A widespread belief in particle physics is that the
relativistic concept of a {\it charged\/} particle does not exist
due to the infra-red structure of gauge theories (see, for example,
the conclusions to Ref.\ \FAD). This point of view is based on the
observation\up{\KIBBLE} that the Green's functions for the charged
matter fields do not have a simple pole like structure but rather a
branch cut,
thus signalling the need for a form of asymptotic
dynamics more complicated than that of the free theory\up{\FAD}.
The interpretation of this result is that in QED  a  soft photon
cloud always surrounds  each physical charge (see the discussion on
page~524 of Ref.\ \JR).  Although there
is now a well developed industry which, at the level of
cross sections, avoids most of
the problems associated with our lack of
knowledge of the form of this asymptotic dressing\up{\CSS},
even in QED it is still not completely
understood how to circumvent these difficulties\up{\CONT-\CONTR}.
The situation in non-abelian theories is
much more complicated as the form of the asymptotic dynamics is
poorly understood\up{\CIAFALONI}.

Here we will show that a relativistic particle
description of charged matter is, in fact, possible. Our
starting point is the observation that one cannot talk about a
physical charge without including some form of gauge dressing,
and it is this {\it combined system\/} that must be identified with a
charged particle. To this end, we
present two equations which determines how gauge degrees of
freedom, such as glue, surround charged matter, such as valence
quarks. We solve these equations for the abelian theory and
demonstrate that the on-shell Green's functions of the solutions
are free of soft divergences and have a pole structure
at all orders in perturbation theory.
This is the first
formulation of QED with this important property.
We conclude with a discussion of the solutions in QCD.

In a gauge theory physical fields must be gauge invariant and, in
particular, charged matter fields\up{\US} are identified with
products of the form $\hm(x)\phi(x)$ where the dressing, $\hm(x)$,
which surrounds the  charged matter, $\phi(x)$, transforms as
$$
\hm(x)\to \hm(x)U(x)\eqn\dressinggt
$$
under the  gauge transformation described by the group element
$U(x)$. Since $\phi$ transforms as $\phi(x)\to U^{-1}(x)\phi(x)$, the
product is gauge invariant.
The explicit $x$-dependence in the charged matter fields
does not imply that this is an operator creating a charge at the
spacetime point $x$. Rather, as discussed in Ref.~\US, such fields
must necessarily be non-local and the $x$-dependence is simply
labelling the
matter field core to the charged field and thus the source of the
dressing. For a particle description of the charged field we need
to investigate whether there is a choice of dressing for which the
field has a well defined momentum.

We will say that the charged field $\hm(x)\phi(x)$ has a sharp
momentum $p^\mu$ if the state created by it is a (generalised)
eigenstate of the momentum operator  $\widehat{P}^\mu=i\pa^\mu$,
with eigenvalue $p^\mu$. Thus the sharpness condition is
$\pa^\mu\bigl(\hm(x)\phi(x)\bigr)=-ip^\mu\hm(x)\phi(x)$.
In terms of the four velocity, $u^\mu$, we can write this sharpness
condition as
$$
u\cd\pa\bigl(\hm(x)\tilde{\phi}(x)\bigr)=0\,,\eqn\sharptilde
$$
where $\tilde{\phi}(x)=e^{imu\cd x}\phi(x)$, which is familiar from
heavy quark effective theory.

We now need to find such sharp
charged states which we can then identify, in an appropriate
asymptotic regime, with the in and out charged particle states of the
theory.
For massive charged matter (which is the only case we will consider
in this paper), the heavy matter sector (or equivalently
the soft gauge sector) is the region where the velocity becomes a
superselection rule\up{\GEORGI} and a particle description
can emerge.  In this sector the matter fields
satisfy the equation
$$
u\cd D(\tilde{\phi}(x))=0\,.\eqn\heavymatter
$$
(In this expression the covariant derivative is taken to be
$D_\mu=\pa_\mu+gA_\mu$ where
$A_\mu$ is the Lie algebra valued (anti-Hermitian) gauge
potential.)
It then follows that  the dressed charge has a sharp
momentum in this heavy sector if the dressing  satisfies the
equation
$$
u\cd\pa(\hm(x))=g\hm(x)u\cd A(x)\,.\eqn\sharpdressing
$$

Equations (\dressinggt) and
(\sharpdressing) are the
fundamental equations for determining the dressing and hence the soft
dynamics of charged particles in gauge theories. The first of these
is a minimal demand: physical charges are
gauge invariant. Previous work on these various solutions, in the
framework of the standard model, was summarised and extended in Ref.\
\US\ (see also Ref.'s \FIRST -\LAST).
The second equation is a new, kinematical requirement which
removes the ambiguity associated with the plethora of solutions to
(\dressinggt).
We now note
that in QED we can explicitly solve {\it both} of these
equations. The resulting dressing can be
shown to factor into the product of two terms:
$$
\hm(x)=e^{-ieK(x)}e^{-ie\chi(x)}\,.\eqn\qedsoln
$$
The $K$-dependent factor is gauge invariant and
describes the response of the
charge to the other charges in the system. The
second term in the dressing represents the soft core of the
dressing and its form was essentially guessed in Ref.\ \US\ through
an analysis of the resulting electric and magnetic fields of the
dressed particle. The explicit forms of these terms are
$$
\eqalign{
K(x)=&-\int_{\Gamma} d\Gamma(\eta+v)^\mu \frac{\pa^\nu
F_{\nu\mu}}{{\cal G}^v\cdot \pa}
\cr
\chi(x)=&\frac{{\cal G}^v_\mu A^\mu}{{\cal
G}^v\cd\pa}(x):=-\frac{\gamma}{4\pi}\int d^3z\frac{{\cal G}^v_\mu
A^\mu(x^0,\underline{z})}
{\|\underline{z}-\underline{x}\|_v}\,.
}\eqn\sixx
$$
In these expressions ${\cal G}^v_\mu=(\eta+v)_\mu(\eta-v)\cdot
\pa-\pa_\mu$ with $\eta=(1,\underline{0})$, a unit timelike
vector; $v$ the orthogonal velocity vector
$v=(0,\underline{v})$ where $\underline{v}$ is the velocity three
vector  of the particle; $\gamma=(1-\underline{v}^2)^{-\frac12}$ and the metric
convention is $(+,-,-,-)$. The contour, $\Gamma$, in the
definition of $K$ is the past world line of a particle moving with four
velocity $u^\mu$. The explicit form of the
inverse to ${\cal G}^v\cd\pa$ is given by
$$
\frac1{\|z\|_v}:=
\frac1{2\pi^2\gamma}\int\!
d^3k\frac{e^{i \underline{k}\cd\underline{z}}}
{V^v\cd k}\,,\eqn\no
$$
where $V^v_\mu=(\eta+v)_\mu(\eta-v)\cdot k-k_\mu$. This is well defined for massive
matter ($|\underline{v}|<1$).  As $v\to0$
we obtain the usual inverse to the three-dimensional Laplacian.
Eq.~\sixx\ makes manifest the spatial non-locality of the physical
charges.

On physical states the $K$ term of the dressing may be reexpressed
in terms of the matter current. As such it is the
analogue of the Coulombic term in infra-red dynamics\up{\JR} and,
as can be verified explicitly,
does not take part in the soft dynamics of  the abelian theory.

In the remainder of this paper we will show that, using the
non-Coulombic, $\chi$-dependent part of the
dressing,
we remove the soft divergences due to virtual photons
in the Green's functions of these dressed charges,
$e^{-ie\chi(x)}\phi(x)$, at all orders in
scalar QED. Note that
there are no collinear divergences as we
assume massive electrons and that we use the scalar theory for
simplicity: the electron's
spin does not affect the infra-red structure. We will consider the
terms in the dressed Green's functions that have a simple pole for
each external leg. Generally an infra-red divergence appears in the
residues of these pole structures when we go on-shell. We
will show that these soft divergences cancel when we include our
dressings.
We start with the propagator of the above (dressed) scalar
electron which at one loop is given by the four diagrams of
Fig.\ 1 (with {\sl
(b)} and {\sl (c)} identical in the scalar theory).
The new Feynman rule for $n$ photons at a
dressing vertex is shown in Fig.\ 2.
We will work in Feynman
gauge but gauge invariance will be always apparent in our final
results. Note that the
dressed fermion propagator is treated in detail in Ref.\ \FAST\ and
the scalar case
in Ref.\ \SCAL: IR-finite results were obtained in both cases and the
renormalisation
constants were calculated. We therefore extract the IR-divergent terms in
the residue of
the pole in the propagator. After some algebra they
are found to be
$$
\hbox{Res}[iS^{{\rm IR}}_{\rm pole}]=
\frac{ie^2}{(2\pi)^4}\int\!d^4k
\( \frac{p_\mu}{p\cdot k} -\frac{V^v_\mu}{V^v\cdot k}\)
\frac{g^{\mu\nu}}{k^2}
\( \frac{p_\nu}{p\cdot k} -\frac{V^v_\nu}{V^v\cdot k}\)
\,,
\eqn\todie
$$
where we are only integrating over soft momenta less than some
cut-off.
Note that the  divergences  in (\todie) are of logarithmic type.
Replacing $g^{\mu\nu}/k^2$ by a more general propagator shows
the gauge invariance of this result. We now need to
show that it in fact vanishes: this may be done either by direct
calculation or by realising that effectively we may write (with
$p^*=(p^0,-{\underline p})$)
$$
\frac{V_\mu^v}{V^v\cdot k}\to
\frac{p_\mu}{p\cdot k}-\frac{m^2\gamma^2k_\mu}{p\cdot k\,p^*\cdot k}\,
\eqn\trick
$$
which holds when we evaluate at the residue of the $k^2$ poles (which
are well known to be the only poles that yield soft divergences) and go
on shell in the correct fashion: i.e., at
$p=m\gamma(1, {\underline v})$. The second
term clearly will not contribute to (\todie) and we see that {\it the
soft
divergences all cancel}.

The pole structure that we obtain has, as a special case, the limit
${\underline p}\to0$, where the dressed charge and the
undressed one coincide in Coulomb gauge. Thus we see that our
construction  explains and extends a result due to
Kibble\up{\KIBBLE}, who noted
that the propagator in the Coulomb gauge has a pole if we are
at the static point on
the mass shell, $p=(m,{\underline 0})$.
We also remark that we may use Sect.\ 3D
of Ref.\ \JACKSOL\ to see that
the dressed propagator is indeed IR-finite at all orders if we choose
the correct point on the mass shell.

We now consider an interaction with a source. We will assume that the
vertex
has the form $m^2\phi^{*}\phi$, which is renormalisation group
invariant.
For simplicity we take a charge initially at
rest which emerges from the interaction with velocity $v$. We now
multiply in a factor of
 $1/{\sqrt{{\zz 0\zz v}}}$, where $\zz v$ is the
infra-red finite\up{\FAST,\SCAL}, $v$-dependent
wave-function  renormalisation constant associated with a particle
moving with three velocity $\underline v$. This infra-red finite factor
will lead to various useful cancellations of diagrams.
As far as the residues of the poles are
concerned their diagrammatic form is shown in Fig.\ 3.
Using this, we obtain the covariant contribution from Fig. 4
while the explicitly non-covariant terms are
expressed by Fig.\ 5. Note
the amputated legs on the last two terms of this figure (which
are each to be
understood as the product of their two parts). Calculating these
latter
diagrams we find for the IR-divergences in the residue the
gauge-invariant
result
$$
-\frac{m^2e^2}2\frac1{p^2_1-m^2}\frac1{p^2_2-m^2}
\int\!\frac{d^4k}{(2\pi)^4}
\(\frac{V_\mu^v}{V^v\cdot k} -
\frac{V_\mu^0}{V^0\cdot k}
\)
\frac{-ig^{\mu\nu}}{k^2}
\(\frac{V_\nu^v}{V^v\cdot k} -
\frac{V_\nu^0}{V^0\cdot k}
\)
\,.
\eqn\ginv
$$
Using (\trick) we may easily see that this cancels the standard
covariant integral which is just (\ginv) with an opposite sign
and the $V$'s replaced by
the
appropriate momentum  $p_i$.
The extension of this argument to other
processes is direct.

To show that this is true at all orders we consider the structure
of the propagator. As well as the usual dressing-independent
diagrams which yield the form, $i/(p^2-m^2-\Sigma)$, we can have
(at either end) a \lq blob\rq\ with lines emanating from a dressing
vertex, this we denote by $\tilde\Sigma$. This then adds to the
propagator the term, $2i\tilde\Sigma/(p^2-m^2-\Sigma)$. A blob at
both ends contributes, ${i{\tilde\Sigma}^2}/(p^2-m^2-\Sigma)$. If
these end blobs overlap, there is no pole contribution to the
residue, except in one important case: when this overlap is solely
due
to lines  directly
linking  one dressing vertex to the other. This is a consequence of
the
Feynman rules for the dressing vertex (see Ref.'s \FAST\ and \SCAL\
for the one loop example).
It may be seen that the IR-divergent
contribution of such \lq rainbow\rq\ lines to the residue
factorise, yielding a factor of $-C_{vv}$ for each line where we
define
$$
C_{vv'}=\intden \frac{V^v\cdot V^{v'}}{(V^v\cdot k)(V^{v'}\cdot
k)k^2}
\,.
\eqn\no
$$
Summing up terms, and noting the $1/n!$ symmetry factor for $n$ such
lines, yields
an exponential.
The general structure of the IR-divergent part of the propagator pole
is thus
$$
\frac{i(1+\tilde\Sigma)^2}{p^2-m^2-\Sigma} \exp{(-C_{vv})}
\,.
\eqn\aop
$$
We may conclude that dressings do not change the mass shift (as
seen\up{\FAST,\SCAL}
in explicit one-loop calculations) and that the wave function
renormalisation constant
factors into a covariant and a non-covariant part, $
Z_2^v=Z_2^{\eightpoint cov}(1+\tilde\Sigma)^2\exp(-C_{vv})$.

Armed with this we consider the vertex at all orders.
The general class of diagrams with two simple poles and possible
IR~divergences is shown in Fig.\ 6,
i.e., the diagrams
where there are possible covariant vertex corrections from
one leg to another (in the leftmost blob),
covariant corrections on the external legs, \lq
end blob\rq\ corrections on the legs and possible rainbow
corrections from one dressing to the other. The use of a black dot,
$\bullet$, here denotes that there may or may not be an \lq end
blob\rq\ ($\tilde\Sigma$) at the ends of one or both of the lines.
Other diagrams are either IR-finite or will not yield both poles. For
example, if a line from, say, the upper  dressing vertex is attached
to a covariant vertex on the lower scalar matter
leg there will not be a
pole associated with the upper leg.

Factorising the rainbow dressings yields the diagrams of Fig.\ 7.
Since there may or may not be an end blob
in these lines, we may write these end factors as $(1+\tilde\Sigma)$.
They therefore {\it cancel} the $\tilde\Sigma$ dependence from each
of
the external leg $(Z_2^v)^{-1/2}$ factors.
Diagrammatically we thus obtain Fig.\ 8
and we see that the dressing effects have exponentiated. Since
the remaining covariant diagrams times $1/{Z_2^{\eightpoint
cov}}$ are known to
exponentiate, we see that {\it all the soft effects exponentiate}
in the residue of the double poles. We have seen that they cancel
at one loop, this now directly implies that  this residue
is IR~finite at all orders.  This
demonstration shows the validity of the dressing equation.
We also note that the
extension of this to other vertices is straightforward: we conclude
that {\it  correctly
dressed  on-shell Green's functions are infra-red finite to all
orders}.
Full details of the proof will be presented elsewhere together with
the details of the one-loop calculations.

We now finish with some remarks on the implications  of this
paper for the non-abelian theory. The equations (\dressinggt)
and (\sharpdressing)
open a way to {\it deriving} how glue dresses quarks.
In Ref.\ \US\ it was shown that there are no global
solutions to the dressing equation (\dressinggt) in an unbroken
non-abelian theory, however perturbative solutions can be
constructed and non-perturbative effects can be incorporated into
them. Perturbative solutions for the gluonic dressing are relevant
for
short distance physics such as initial jet formation. At larger
scales
non-perturbative effects will dominate.
The sharpness condition (\sharpdressing) on the dressing
leads to sharp quark states in the region where the heavy mass
equation (\heavymatter) holds. For heavy quarks this will be at a
scale significantly greater than $\Lambda_{{\rm QCD}}$ and thus for
such quarks a non-abelian extension of the abelian dressing
described above would be expected within a Gribov horizon where the
dressing equation can be solved. For the lighter quarks, their
heavy sector is at a scale lower than $\Lambda_{{\rm QCD}}$ and
thus non-perturbative contributions to the dressing, such as
condensates\up{\US}, would be expected to play a dominant role even
before the effects of the horizon become manifest.
We note here
that the inter-quark potential can be found rather directly from
dressings (see Sect.\ 7 of Ref.\ \US\ and Ref.\ \HAJO) and further that it has been
argued\up{\ZWAN} that the incorporation of non-perturbative effects
associated with the Gribov ambiguity, and hence with the breakdown of
solutions to (\dressinggt), lead to a linearly rising potential. The
incorporation of non-perturbative effects in dressings which solve
(\dressinggt) and (\sharpdressing) and their impact on the potential
will be studied elsewhere.

\bigskip
\ni {\bf Acknowledgements:} EB received support from CICYT
research project AEN95-0815.
MJL thanks the Universitat Aut\`onoma de Barcelona for their
hospitality while much of this work was carried out.

\bigskip
\bigskip
\bigskip

  \vfill\eject\immediate\closeout\reffile
  \centerline{{\bf References}}\bigskip\frenchspacing%
  \input refs.tmp
\vfil\eject

\ni {\bf The Figures}
\bigskip\bigskip
\fig{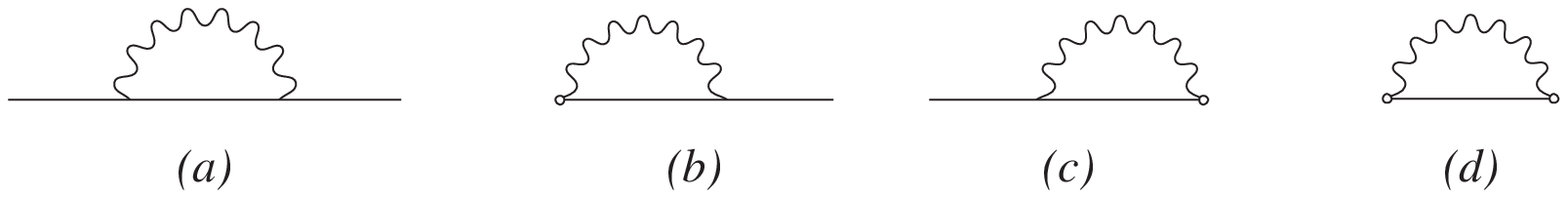}{12cm}
\ni {\bf Fig.\ 1:} The one-loop propagator for the dressed field,
$e^{-ie\chi}\phi$.
\bigskip
\centerline{
\hbox{
\fig{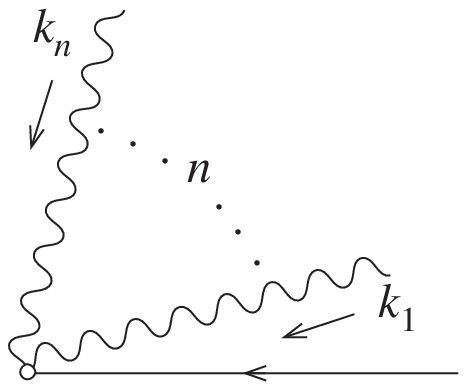}{2.0cm}
    \vbox{\hbox{$\displaystyle = \frac{eV^1_{\mu_1}}{V^1\cdot k^1}
\cdots  \frac{eV^n_{\mu_n}}{V^n\cdot k^n}
$}\vskip.3cm
}}}
\medskip\ni
\ni {\bf Fig.\ 2:} The Feynman rule for the dressing vertex.
\bigskip
\fig{z2.eps}{12cm}
\ni {\bf Fig.\ 3:} Diagrammatic expression for $Z_2$.
\bigskip

\fig{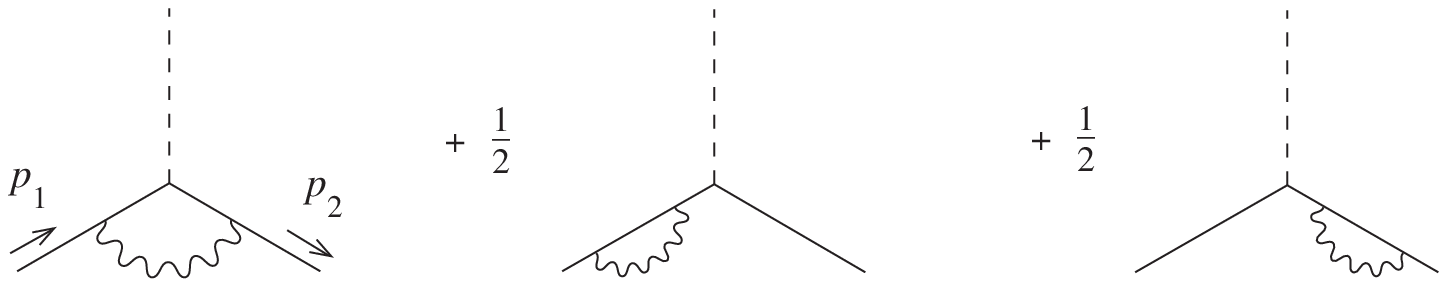}{10cm}
\ni {\bf Fig.\ 4:} Covariant part of the vertex.
\bigskip
\vfil\eject
\fig{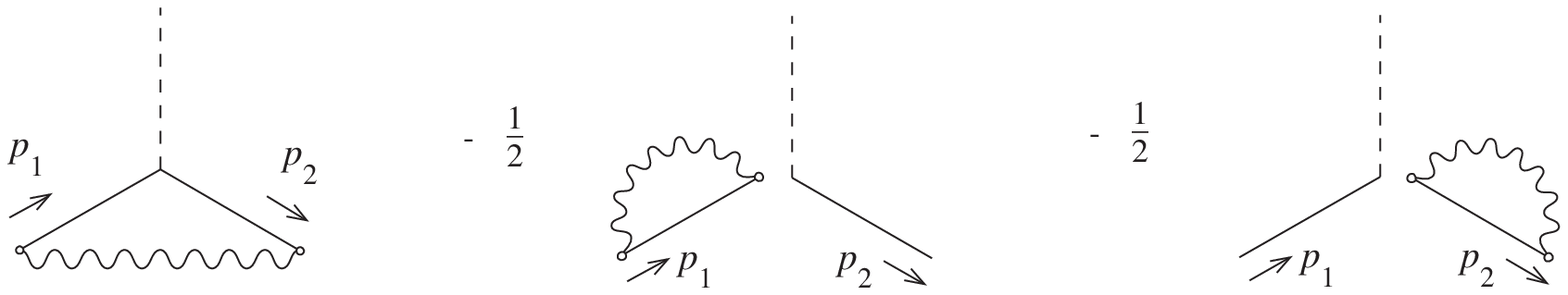}{10cm}
\ni {\bf Fig.\ 5:} Non-covariant part of the vertex.
\bigskip

\centerline{
\hbox{
\fig{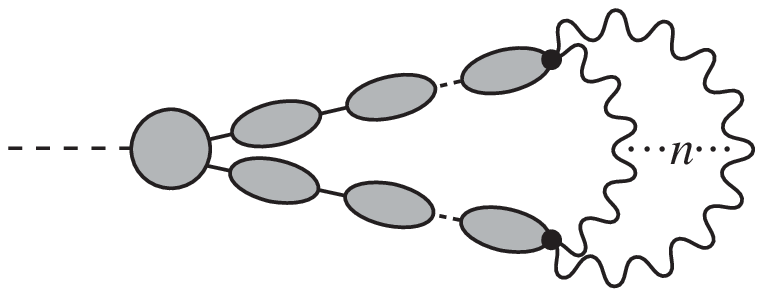}{5.0cm}
\vbox{\hbox{$\displaystyle\frac1{\sqrt{Z_2^v Z_2^{v'}}}$
}\vskip.5cm }}}
\medskip
\ni {\bf Fig.\ 6:} Diagrams with both poles and IR divergences.

\bigskip

\centerline{
\hbox{
\fig{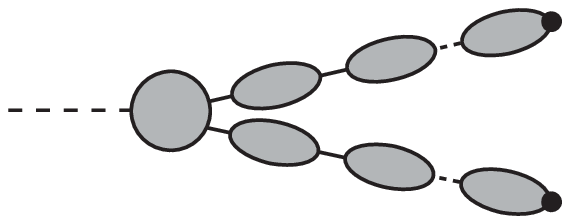}{4.0cm}
\vbox{\hbox{$\;\displaystyle \exp\(-C_{vv'}\)
\frac1{\sqrt{Z_2^vZ_2^{v'}}}$
}\vskip.3cm
}}}
\ni {\bf Fig.\ 7:} The preceding figure after factorisation.

\bigskip

\centerline{
\hbox{
\fig{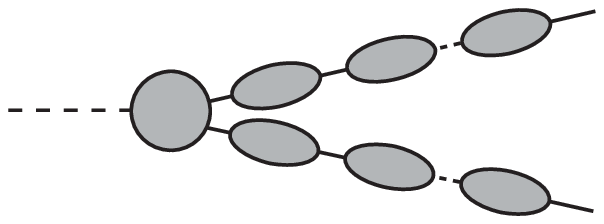}{4.0cm}
\vbox{\hbox{$\;\displaystyle
\exp\({-C_{vv'}+\frac12C_{vv}+\frac12C_{v'v'}}\)
\frac1{Z_2^{\eightpoint cov}}
$
}\vskip.3cm
}}}
\ni {\bf Fig.\ 8:} Exponentiation of the dressings.

\bye